\documentclass{article}
\usepackage{float}
\usepackage{authblk}
\usepackage{algorithm}
\usepackage{algorithmicx}  
\usepackage{algpseudocode}  
\usepackage{setspace}
\usepackage{bm}
\usepackage[draft]{hyperref}
\usepackage{url}
\usepackage{amsmath}
\usepackage{stfloats}
\usepackage{mathrsfs}
\usepackage{dcolumn}
\usepackage{subfigure}  
\usepackage{multirow}
\usepackage{graphics}
\usepackage{graphicx}
\usepackage{geometry}
\usepackage{amsfonts} 
\usepackage{amssymb}
\usepackage{amsthm}
\usepackage{longtable}
\usepackage{booktabs}
\title{\textbf{A Novel Unified Framework for Solving Reachability, Viability and Invariance Problems}}

\author[a,b]{Wei Liao}
\author[a,b]{Taotao Liang\thanks{Wei Liao and Taotao Liang contributed equally to this work.}}
\author[a,b]{Xiaohui Wei \thanks{Corresponding author: wei\_xiaohui@nuaa.edu.cn}}
\author[c]{Jizhou Lai}
\affil[a]{\footnotesize{ Key laboratory of Fundamental Science for National Defense-Advanced Design Technology of Flight Vehicle, 
Nanjing University of Aeronautics and Astronautics, Nanjing, Jiangsu, China }}
\affil[b]{\footnotesize{State Key Laboratory of Mechanics and Control of Mechanical Structures, 
Nanjing University of Aeronautics and Astronautics, Nanjing, Jiangsu, China}}
\affil[c]{\footnotesize{Navigation Research Center, Nanjing University of Aeronautics and Astronautics, Nanjing, Jiangsu, China}}

\newtheorem{remark}{\textbf{Remark}}
\newtheorem{thm}{\textbf{Theorem}}

\newtheorem{definition}{\textbf{Definition}}
\begin{document}
\maketitle
\begin{abstract}
	The level set method is a widely used tool for solving reachability and invariance problems. 
	However, some shortcomings, such as the difficulties of handling dissipation function and constructing 
	terminal conditions for solving the Hamilton-Jacobi partial differential equation, 
	limit the application of the level set method in some problems with non-affine nonlinear systems and irregular target sets. 
	This paper proposes a method that can effectively avoid the above tricky issues and thus has better generality. 
	In the proposed method, the reachable or invariant sets with different time horizons are characterized by some non-zero sublevel 
	sets of a value function. This value function is not obtained by solving a viscosity solution of the partial differential 
	equation but by recursion and interpolation approximation. At the end of this paper, 
	some examples are taken to illustrate the accuracy and generality of the proposed method.
\end{abstract}
\quad \\
\textbf{Keywards:} Reachability, Viability, Invariance, Optimal Control, Nonlinear system.

\section{Introduction}
%1 Polynomial Level-Set Method for Polynomial System Reachable Set Estimation
%2 Hamilton-Jacobi Reachability: A Brief Overview and Recent Advances
%3 Under-Approximating Reach Sets for Polynomial Continuous Systems
% 别忘记把t 和t_f 的问题缕一缕

%1 Polynomial Level-Set Method for Polynomial System Reachable Set Estimation
%2 Hamilton-Jacobi Reachability: A Brief Overview and Recent Advances
%3 Under-Approximating Reach Sets for Polynomial Continuous Systems
% 别忘记把t 和t_f 的问题缕一缕
Many strict definitions, such as controllability and observability, 
and many mature analytical methods have been proposed to analyze linear control systems' behavior (\cite{b1}). 
These methods have been widely applied in engineering for solving stability and reliability problems. 
However, in nonlinear control systems, properties similar to those mentioned above are difficult to define and analyze (\cite{b2}).

Nonlinear control systems' behavior can be studied using reachability and invariance analyses. 
In reachability and invariance analysis, it is first necessary to specify a time horizon and a target set in the state space, 
and then to compute the reachable and invariant sets of the target set (\cite{a001,frank2}). 
Computing these sets, on the other hand, is a difficult task. 
Because of the variety of system states and control inputs, checking each state one at a time takes a long time (\cite{a002,i1}). As a result, 
simulation-based approaches are insufficient, and formal verification methods are required. Unfortunately, with the exception of 
a few solvable problems (\cite{a003}), computing these sets exactly is often intractable. 
Therefore, it's customary to compute their approximate expressions.

In recent decades, a variety of methods have been presented, which can be split into two categories: 
Lagrangian methods (\cite{a004}) and state space discretization methods. Lagrangian methods, 
which can handle high-dimensional problems but are mainly limited to linear dynamic systems, 
include ellipsoidal methods (\cite{i3,a006,i4}), polyhedral methods (\cite{frank1,a005,a007}), and support vector machines (\cite{a004}). 
The state space discretization-based techniques 
require less of the form of the dynamical system and can solve non-linear problems.
The level set method (\cite{a001,a002,a008,ijc2}) and the distance fields over grids (DFOG) method (\cite{a0010,a0011,a0012}) 
are the two most common methods in this category, 
with the level set method being more widely applied in engineering.

More importantly, the level set method creatively characterizes the reachable and invariant sets in a unified form.
In the level set method, the reachable or invariant set is represented as a zero level set of a 
value function, which is the viscosity solution of a Hamilton-Jacobi (HJ) partial differential equation (PDE) 
\textit{without} running cost function (\cite{a001}).
Different types of sets can be obtained by setting different terminal conditions and switching the MIN or MAX 
operator during the computation.
On this basis, several mature toolboxes have been developed (\cite{tr1,web1,web2,web3}), and solved a 
variety of practical engineering problems, 
such as flight control systems (\cite{i5,a009,a0013}), ground traffic systems (\cite{i6,a0014}), 
and air traffic management systems (\cite{a0015}).

While the level set approach has evolved into a well-developed and dependable tool over time, it still has some limitations.

\begin{itemize}
	\item [(1)] The level set method entails the viscosity solution of an HJ PDE, which requires addressing an optimization problem
	known as the dissipation function. To our best knowledge, although this issue can be 
	simplified in affine nonlinear systems, there is no practical approach for solving it in more general systems (\cite{tr1,a0015}). 
	Therefore, the application of level set methods is mainly limited to affine nonlinear systems. 
	\item [(2)] The terminal condition of the HJ PDE is always set as a signed 
	distance function of the target set. However, it is difficult to construct such a function for an irregular target set.
	In some studies, the irregular target set is replaced by a rectangle when constructing the required signed distance function (\cite{a009,a0013}).
	\item [(3)]   {In the level set method, 
	saving the reachable or invariant sets for different time horizons requires saving the solutions of the HJ PDEs at different time points, 
	which makes the storage space consumption proportional to the number of reachable or invariant sets to be saved.}
\end{itemize}

To overcome these shortcomings, this paper proposes a novel unified framework for solving reachability
and invariance problems. In this framework, the reachable and invariant sets 
can be described as some non-zero sublevel sets of some value functions, which  
are solutions of some HJ PDEs \textit{with} running cost functions, and the HJ PDEs are 
solved by recursion and interpolation.
Such a mechanism has the following advantages:
\begin{itemize}
	\item[(1)] Since there is no PDE solution involved, the computation of the dissipation function is effectively avoided. 
	Therefore, the proposed method can be applied to non-affine nonlinear systems.
	\item[(2)] As it is unnecessary to construct a signed distance function of the target set as the terminal condition of the HJ PDE, 
	this method can easily handle irregular target sets.
	\item[(3)]   {The reachable or invariant sets for different time horizons can be characterized by the solutions of the HJ 
	PDEs at the same time point. Therefore, the required storage space is independent of 
	the number of reachable or invariant sets to be saved.}
\end{itemize}

The structure of this paper is as follows. Section 2 introduces reachable, 
viable, and invariant sets and briefly describes the level set method. 
Section 3 describes the proposed method in detail. In Section 4, 
a two-dimensional problem is taken as an example to analyze the accuracy of the proposed method. 
Section 5 takes an engineering problem as an example to 
illustrate the generality of our method. Finally, a brief conclusion is presented in Section 6.
% \section*{Acknowledgements}
% \par The authors gratefully acknowledge support from National Defense Outstanding Youth Science Foundation (Grant No. 2018-JCJQ-ZQ-053), 
% and Central University Basic Scientific Research Operating Expenses Special Fund Project Support (Grant No. NF2018001).
% Also, the authors would like to thank
% the anonymous reviewers, associate editor, and editor for
% their valuable and constructive comments and suggestions.

% \bibliographystyle{Bibliography/IEEEtranTIE}
% % \bibliography{Bibliography/IEEEabrv,Bibliography/BIB_xx-TIE-xxxx}\ %IEEEabrv instead of IEEEfull
% \bibliography{mybib}

\section{Preliminaries}
\subsection{Statement of Problem}
Consider a continuous time control system with fully observable state:
\begin{align}
	\label{sys1}
	\dot{s}=f(s,u)
\end{align}
where $s\in \mathbb{R}^n$ is the system state, $u\in\mathcal{U}$ is the control input ($\mathcal{U}$ is a closed set).
The function $f(.,.):\mathbb{R}^n\times \mathcal{U}\to \mathbb{R}^n$ is Lipschitz continuous and bounded.
Let $\mathscr{U}$ denote the set of Lebesgue measurable functions from the time interval $[0,\infty)$ to $\mathcal{U}$.
Then, given the initial state $s_0$ at time $t_0$ and $u(.)\in \mathscr{U}$, the evolution of system (\ref{sys1}) in time
interval $[t_0,t_1]$ can be expressed as a continuous trajectory $\phi_{t_0}^{t_1}(.,s_0,u):[t_0,t_1]\to \mathbb{R}^n$ 
and $\phi_{t_0}^{t_1}(t_0,s_0,u)=s_0$. Given a target set $K$ and a time horizon $T$, 
four important definitions can be proposed:
\begin{definition}
	\textbf{Maximal reachable set} 
	\begin{align}
		\begin{split}
		&\mathcal{R}_{\mathrm{max}}(K,T)
		=\left\{s_0| \exists u(.)\in\mathscr{U}\ \exists t\in [0,T]\ \phi_{0}^{T}(t,s_0,u)\in K  \right\}
		\end{split}
	\end{align}
\end{definition}

\begin{definition}
	\textbf{Minimal reachable set} 
	\begin{align}
		\begin{split}
		&\mathcal{R}_{\mathrm{min}}(K,T)
		=\left\{s_0| \forall u(.)\in\mathscr{U}\ \exists t\in [0,T]\ \phi_{0}^{T}(t,s_0,u)\in K  \right\}
		\end{split}
	\end{align}
\end{definition}

\begin{definition}
	\textbf{Maximal invariant set} 
	\begin{align}
		\begin{split}
		&\mathcal{I}_{\text{max}}(K,T)
		=\left\{s_0| \exists u(.)\in\mathscr{U}\ \forall t\in [0,T]\ \phi_{0}^{T}(t,s_0,u)\in K  \right\}
		\end{split}
	\end{align}
\end{definition}

\begin{definition}
	\textbf{Minimal invariant set} 
	\begin{align}
		\begin{split}
		&\mathcal{I}_{\text{min}}(K,T)
		=\left\{s_0| \forall u(.)\in\mathscr{U}\ \forall t\in [0,T]\ \phi_{0}^{T}(t,s_0,u)\in K  \right\}
		\end{split}
	\end{align}
\end{definition}

The keys to solve the reachability and invariance problems are to 
compute the above-mentioned sets.

\subsection{Level set Method}
In the level set method, two value functions $V_1$ and $V_2$ are characterized as viscosity solutions to the following 
Hamilton-Jacobi (HJ) partial differential equations (PDEs):
\begin{align}
	\label{lsHJeq}
	\begin{split}
		&\left\{\begin{array}{l}
		\displaystyle{ \frac{\partial V_1}{\partial t}(s,t)+\min \left[0,\max_{u\in \mathcal{U}} \frac{\partial V_1}{\partial s}(s,t) f(s,u) \right]=0}\\
		\text{s.t.  } V_1(s,T)=l(s)
		\end{array}\right.\\
		&\left\{\begin{array}{l}
			\displaystyle{ \frac{\partial V_2}{\partial t}(s,t)+\min \left[0,\min_{u\in \mathcal{U}} \frac{\partial V_2}{\partial s}(s,t) f(s,u) \right]=0}\\
			\text{s.t.  } V_2(s,T)=l(s)
		\end{array}\right.
	\end{split}
\end{align}
where $l(.):\mathbb{R}^n \to \mathbb{R}$ is a Lipschitz continuous function and satisfies:
\begin{align}
	K=\left\{s\in\mathbb{R}^n|l(s)\geq 0 \right\}
\end{align}
A common choice for the function $l(.)$ is the signed distance to the target set $K$.
Then, the maximal and minimal invariant sets are denoted as zero super-level sets of two value functions, i.e.:
\begin{align}
	\begin{split}
		\mathcal{I}_{\text{max}}(K,T)=\left\{s_0| V_1(s_0,0) \geq 0 \right\}\\
		\mathcal{I}_{\text{min}}(K,T)=\left\{s_0| V_2(s_0,0) \geq 0 \right\}
	\end{split}
\end{align}
Some literature point out that the invariance and reachability problems
are duals of one another (\cite{a001}), i.e.
\begin{align}
	\label{rlt1}
	\begin{split}
		\mathcal{R}_{\text{max}}(K,T)=\complement_{\mathbb{R}^n} \left[\mathcal{I}_{\text{min}} ({\complement_{\mathbb{R}^n } K },T)\right] \\
		\mathcal{R}_{\text{min}}(K,T)=\complement_{\mathbb{R}^n} \left[\mathcal{I}_{\text{max}} ({\complement_{\mathbb{R}^n } K },T)\right] 
	\end{split}
\end{align}
% Thus, the level set method represents the reachable, viable and invariant sets in a unified form.
% Based on this principle, some mature toolboxes (\cite{web1,a0015}) have been developed and have greatly promoted the development of 
% computational techniques for reachable and invariant sets and solved many practical engineering problems, 
% such as safe flight envelope (\cite{i5,a009,a0013}), ground traffic management systems (\cite{i6,a0014}), air
% traffic management systems (\cite{a0015}), etc.

The user of the level set toolboxes is required to provide the following parameters:
a computational domain, the number of grid nodes in each dimension of the computational domain, the terminal condition of the HJ PDE, 
the Hamiltonian, and the dissipation function.

\begin{remark}
	In the level set method,
	for $T_1,...,T_M \in [0,\infty)$, the invariant sets of these time horizons are characterized by the 
	zero super-level sets of the solutions of Eq. (\ref{lsHJeq}) at time points $T_1,...,T_M$, i.e.
	\begin{align}
		\begin{split}
			&\mathcal{I}_{\text{max}}(K,T_1)=\left\{s_0| V_1(s_0,T-T_1) \geq 0 \right\} \ \ \ \mathcal{I}_{\text{min}}(K,T_1)=\left\{s_0| V_2(s_0,T-T_1) \geq 0 \right\} \\
			&...\\
			&\mathcal{I}_{\text{max}}(K,T_M)=\left\{s_0| V_1(s_0,T-T_M) \geq 0 \right\} \ \ \ \mathcal{I}_{\text{min}}(K,T_1)=\left\{s_0| V_2(s_0,T-T_M) \geq 0 \right\} 
		\end{split}
	\end{align}
	and the solutions of Eq. (\ref{lsHJeq}) at different time are generally different from each other.
	This means that the storage space required to save these invariant sets is proportional to $M$.
\end{remark}

\section{Recursive Value Function Method}
In this section, we propose a method that can avoid the above-mentioned difficulties, 
namely recursive value function method.
% We first introduce a definition, minimal reachable set \cite{a002}, before formally presenting this method. 
% \begin{definition}
% 	\textbf{Minimal reachable set} 
% 	\begin{align}
% 		\begin{split}
% 		&\mathcal{R}_{\mathrm{min}}(K,T)\\
% 		&=\left\{s_0| \forall u(.)\in\mathscr{U}\ \exists t\in [0,T]\ \phi_{0}^{T}(t,s_0,u)\in K  \right\}
% 		\end{split}
% 	\end{align}
% \end{definition}
% Informally, $\mathcal{R}_{\mathrm{min}}(K,T)$ is a set of states from which trajectories can reach the target set 
% for any $u(.)\in \mathscr{U}$ within time $T$. 
% Similar to the relationship between the maximal reachable set and the invariant set, 
% the viable set and the minimal reachable set are also dual to each other, i.e.:
% \begin{align}
% 	\label{rlt2}
% 	\mathcal{V}(K,T)=\complement_{\mathbb{R}^n} \left[\mathcal{R}_{\mathrm{min}} \left(\complement_{\mathbb{R}^n} K,T\right)\right] 
% \end{align}
According to Eq. (\ref{rlt1}), the maximal and minimal invariant sets can be obtained by computing the minimal 
and maximal reachable sets, respectively, 
% while the viable set can be represented by computing the minimum reachable set. 
To avoid repetitive descriptions, this paper focuses on the computation of the maximal and minimal reachable sets.

\subsection{Characterizations of minimal and maximal reachable sets}
Consider a case where the control input $u(.)$ is chosen to avoid the trajectory from reaching the
target set or delay the time when the trajectory first touches the target set.
With such a control input, the initial state of a trajectory that can touch the target set within the time horizon $T$
must be an element of the minimal reachable set. 
Thus, we can construct the following value function:

\begin{align}
	\label{maxvalue}
	W_1(s_0)=\begin{cases}
		0,&s_0\in K\\
		\left\{\begin{array}{rl}
			\displaystyle{\max_{u(.)}} &t_f\\
			\text{s.t.}& \dot{s}(t)=f(s(t),u(t))\ \forall t\in [0,t_f]\\
				 \ &s(0)=s_0\\
				 \ &u(t)\in\mathcal{U} \ \forall t\in [0,t_f]\\
				 \ &s(t)\notin K \ \forall t\in [0,t_f)  
			\end{array}\right\},   & \begin{array}{l} s_0\notin K\text{ and }\forall u(.)\in \mathscr{U}\ \\  \exists t\in [0,\infty)\ \phi_{0}^{\infty}(t,s_0,u)\in K \end{array}\\
		\infty,  &\text{otherwise}
	\end{cases}
\end{align}

Then the minimal reachable set is the $T-$sublevel set of function $W_1(.)$, i.e.:
\begin{align}
	\mathcal{R}_{\mathrm{min}}(K,T)=\left\{s_0|W_1(s_0)\leq T  \right\}
\end{align}

Consider another case in which the control input $u(.)$ is aims to steer the system state 
to reach the target set in the shortest possible time. 
If a trajectory can enter the target set within time horizon $T$ in this case, 
then its initial state must be an element of the maximal reachable set.
Thus, a value function $W_2(.):\mathbb{R}^n \to \mathbb{R}$ can be constructed as Eq. (\ref{minvalue}).

\begin{align}
		\label{minvalue}
		W_2(s_0)=\begin{cases}
			\left\{\begin{array}{rl}
				\displaystyle{\min_{u(.)}} &t_f\\
				\text{s.t.}& \dot{s}(t)=f(s(t),u(t))\ \forall t\in [0,t_f]\\
				 \ &s(0)=s_0\\
				 \ &u(t)\in\mathcal{U} \ \forall t\in [0,t_f]\\
				 \ &s(t_f)\in K
			\end{array}\right\},  & \begin{array}{l}
				\exists u(.)\in \mathscr{U}\  \exists t\in [0,\infty)\\ \phi_{0}^{\infty}(t,s_0,u)\in K
			\end{array}
                     \\
		\infty, & \text{otherwise}
		\end{cases}
\end{align}

Then the maximal reachable set can be characterized by:
\begin{align}
	\mathcal{R}_{\mathrm{max}}(K,T)=\left\{s_0|W_2(s_0)\leq T  \right\}
\end{align}

% Eqs. (\ref{maxvalue}) and (\ref{minvalue}) 
% are optimal control problems with terminal state constraints. 
In order to transform Eqs. (\ref{maxvalue}) and (\ref{minvalue}) to cost-to-go functions
to construct the recursive formulas,
we need to build a modified dynamic system:
\begin{align}
	\label{sys2}
	\dot{s}= \bar{f}(s,u)=\begin{cases}
		f(s,u), &s\notin K\\
		\mathbf{0}, &s\in K
	\end{cases}
\end{align}
and a modified running cost function:
\begin{align}
	\label{runningcost}
	\bar{c}(s) =\begin{cases}
		1, &s\notin K\\
		0, &s\in K
	\end{cases}
\end{align}
Given the state $s_0$ at time $t_0$ and $u(.)\in \mathscr{U}$, the evolution of the modified system (\ref{sys2}) in 
time interval $[t_0,t_1]$
can also be expressed as a continuous trajectory $\bar{\phi}_{t_0}^{t_1}(.,s_0,u):[t_0,t_1] \to \mathbb{R}^n$.

\begin{remark}
	\label{rmk2}
	As long as the trajectory $\bar{\phi}_{t_0}^{t_1}(.,s_0,u)$ evolves outside the target set $K$, 
	it is the same as trajectory $\phi_{t_0}^{t_1}(.,s_0,u)$ and the modified running cost is identically equal to 1.
	When the trajectory $\bar{\phi}_{t_0}^{t_1}(.,s_0,u)$ 
	touches the border of $K$, then it stays on the border under the dynamics (\ref{sys2}) and the 
	running cost is identically equal to 0.
\end{remark}

Then, two cost-to-go functions can be defined:
\begin{align}
	\overline{W}_1(s_0,\bar{T})=\left\{\begin{array}{rl}
	\displaystyle{\max_{u(.)}} & \displaystyle{ \int_0^{\bar{T}} \bar{c}(s(t)) \ dt}\\
	\text{s.t.} & \dot{s}(t)=\bar{f}(s(t),u(t)) \ \forall t \in [0,{\bar{T}}] \\
	\   & s(0)=s_0 \\
	\   & u(t)\in\mathcal{U} \ \forall t\in [0,{\bar{T}}]
	\end{array}\right.
\end{align}

\begin{align}
	\overline{W}_2(s_0,\bar{T})=\left\{\begin{array}{rl}
	\displaystyle{\min_{u(.)}} & \displaystyle{ \int_0^{\bar{T}} \bar{c}(s(t)) \ dt}\\
	\text{s.t.} & \dot{s}(t)=\bar{f}(s(t),u(t)) \ \forall t \in [0,{\bar{T}}] \\
	\   & s(0)=s_0 \\
	\   & u(t)\in\mathcal{U} \ \forall t\in [0,{\bar{T}}]
	\end{array}\right.
\end{align}

Then, we may deduce the following results:
\begin{thm}
	If $0< {\tau} <\bar{T}$, then $\overline{W}_1(s,\bar{T})=W_1(s)$ holds for
	any $s\in \{s_0|\overline{W}_1(s_0,\bar{T})\leq \tau \}$.
\end{thm} 

\begin{proof}
	Case 1: $s\in K$. According to Eq. (\ref{maxvalue}), when $s\in K$, $W_1(s)=0$, 
	and the trajectory of the modified system (\ref{sys2}) initialized from $s$ always stays at $s$.
	Therefore,
	\begin{align}
		\overline{W}_1(s,\bar{T})=\int_0^{\bar{T}}0 dt=0
	\end{align}
	This means that
	\begin{align}
		s\in K\Longrightarrow s\in\left\{s_0| \overline{W}_1(s_0,\bar{T})\leq {\tau} \right\}
	\end{align}
	and
	\begin{align}
		\overline{W}_1(s,\bar{T})=W_1(s)
	\end{align} 

	\noindent
	Case 2: $s\notin K$. In this case, $\overline{W}_1(s,\bar{T})$ satisfies the following inequality:
	\begin{align}
		\overline{W}_1(s,\bar{T})\leq \int_0^{\bar{T}} \max_{s'\in\mathbb{R}^n} \bar{c}(s')\ dt=\int_0^{\bar{T}} 1 dt={\bar{T}}
	\end{align}
	Moreover, $\overline{W}_1(s,\bar{T})=\bar{T}$ if and only if the trajectory of the modified
	system (\ref{sys2}) initialized from $s$ always evolves outside the target set $K$ on the time interval $[0,\bar{T}]$ 
	under any $u(.)\in\mathscr{U}$. Thus, 
	\begin{align}
		\overline{W}_1(s,\bar{T})\leq \tau <\bar{T} 
		\Longrightarrow \forall u(.) \in \mathscr{U} \ \exists t_f\in [0,\bar{T})\ \bar{\phi}_0^{\bar{T}}(t_f,s,u) \in K
	\end{align} 
	Again, according to Remark \ref{rmk2}, $\bar{\phi}_0^{\bar{T}}(.,s,u)$ is the same as 
	${\phi}_0^{\bar{T}}(.,s,u)$ as long as it evolves outside the target set, therefore, 
	\begin{align}
		\overline{W}_1(s,\bar{T})\leq \tau <\bar{T} 
		\Longrightarrow \forall u(.) \in \mathscr{U} \ \exists t_f\in [0,\bar{T})\ {\phi}_0^{\bar{T}}(t_f,s,u) \in K
	\end{align} 
	The preceding equation belongs to the second case in Eq. (\ref{maxvalue}). Consequently,
	\begin{align}
		\begin{split}
			&\overline{W}_1(s,\bar{T})\leq \tau <\bar{T} \Longrightarrow\\
            &\overline{W}_1(s_0,\bar{T})
			=\left\{
            \begin{array}{rl}
                \displaystyle{\max_{u(.)\in \mathscr{U}}} & \displaystyle{\left(\int_0^{t_f}1 dt+\int_{t_f}^{\bar{T}}0 dt\right)}\\
                \text{s.t.}& \dot{s}(t)=f(s(t),u(t))\text{ and } s(t)\notin K\ \forall t\in [0,t_f)\\
						 \ & {s}(t)\in K \ \forall t\in [t_f,\bar{T}]\\
                         \ &s(0)=s_0\\
                         \ &u(t)\in\mathcal{U} \ \forall t\in [0,t_f]
            \end{array} \right.\\
			&=\left\{\begin{array}{rl}
                    \displaystyle{\max_{u(.)}} &t_f\\
                    \text{s.t.}& \dot{s}(t)=f(s(t),u(t))\ \forall t\in [0,t_f]\\
                         \ &s(0)=s_0\\
                         \ &u(t)\in\mathcal{U} \ \forall t\in [0,t_f]\\
                         \ &s(t)\notin K \ \forall t\in [0,t_f)  
                    \end{array}\right.=W_1(s_0)
        \end{split}
	\end{align}
	To sum up, if $\tau<\bar{T}$, $\overline{W}_1(s,\bar{T})=W_1(s)$ holds for
	any $s\in \{s_0|\overline{W}_1(s_0,\bar{T})\leq \tau \}$.
\end{proof}

\begin{thm}
	If $0< {\tau} <\bar{T}$, then $\overline{W}_2(s,\bar{T})=W_2(s)$ holds for
	any $s\in \{s_0|\overline{W}_2(s_0,\bar{T})\leq \tau \}$.
	% For any ${\tau}<\bar{T}$ and any $s\in\left\{s_0| \overline{W}^{\bar{T}}_2(s_0)\leq {\tau} \right\}$, 
	% we have $\overline{W}^{\bar{T}}_2(s)=W_2(s)$.
\end{thm} 

\begin{proof}
	Case 1: $s\in K$. In this case, 
	the trajectory initialized from $s$ touches the target set at time $0$ regardless of the choice of
	the control input, and obviously $W_2(s)=0$.
	The trajectory of the modified system (\ref{sys2}) initialized from $s$ always stays at $s$.
	Therefore,
	\begin{align}
		\overline{W}_2(s,\bar{T})=\int_0^{\bar{T}}0 dt=0
	\end{align}
	This means that
	\begin{align}
		s\in K\Longrightarrow s\in\left\{s_0| \overline{W}_2(s_0,\bar{T})\leq {\tau} \right\}
	\end{align}
	and
	\begin{align}
		\overline{W}_2(s,\bar{T})=W_2(s)
	\end{align} 

	\noindent
	Case 2: $s\notin K$. In this case, $\overline{W}_2(s,\bar{T})$ satisfies the following inequality:
	\begin{align}
		\overline{W}_2(s,\bar{T})\leq \int_0^{\bar{T}} \max_{s'\in\mathbb{R}^n} \bar{c}(s')\ dt=\int_0^{\bar{T}} 1 dt={\bar{T}}
	\end{align}
	Moreover, $\overline{W}_2(s,\bar{T})=\bar{T}$ if and only if there exists a $u(.)\in\mathscr{U}$ 
	and a $t_f\in [0,\bar{T})$ such that the trajectory of the modified
	system (\ref{sys2}) initialized from $s$ can touch the target set $K$ at time $t_f$. 
	Thus, 
	\begin{align}
		\overline{W}_2(s,\bar{T})\leq \tau <\bar{T} 
		\Longrightarrow \exists u(.) \in \mathscr{U} \ \exists t_f\in [0,\bar{T})\ \bar{\phi}_0^{\bar{T}}(t_f,s,u) \in K
	\end{align} 
	Again, according to Remark \ref{rmk2}, $\bar{\phi}_0^{\bar{T}}(.,s,u)$ is the same as 
	${\phi}_0^{\bar{T}}(.,s,u)$ as long as it evolves outside the target set, therefore, 
	\begin{align}
		\overline{W}_2(s,\bar{T})\leq \tau <\bar{T} 
		\Longrightarrow \exists u(.) \in \mathscr{U} \ \exists t_f\in [0,\bar{T})\ {\phi}_0^{\bar{T}}(t_f,s,u) \in K
	\end{align} 
	The preceding equation belongs to the first case in Eq. (\ref{minvalue}). Consequently,
	\begin{align}
		\begin{split}
            \overline{W}_2(s_0,\bar{T})
			&=\left\{
            \begin{array}{rl}
                \displaystyle{\min_{u(.)\in \mathscr{U}}} & \displaystyle{\left(\int_0^{t_f}1 dt+\int_{t_f}^{\bar{T}}0 dt\right)}\\
                \text{s.t.}& \dot{s}(t)={f}(s(t),u(t)) \text{ and } s(t)\notin K  \ \forall t\in [0,t_f)\\
						 \ & s(t)\in K \ \forall t\in [t_f,\bar{T}]\\
                         \ &s(0)=s_0\\
                         \ &u(t)\in\mathcal{U} \ \forall t\in [0,t_f]
            \end{array} \right.\\
			&=\left\{\begin{array}{rl}
                    \displaystyle{\min_{u(.)}} &t_f\\
                    \text{s.t.}& \dot{s}(t)=f(s(t),u(t))\ \forall t\in [0,t_f]\\
                         \ &s(0)=s_0\\
                         \ &u(t)\in\mathcal{U} \ \forall t\in [0,t_f]\\
                         \ &s(t)\notin K \ \forall t\in [0,t_f)  
                    \end{array}\right.=W_2(s_0)
        \end{split}
	\end{align}
	To sum up, if $\tau<\bar{T}$, $\overline{W}_2(s,\bar{T})=W_2(s)$ holds for
	any $s\in \{s_0|\overline{W}_2(s_0,\bar{T})\leq \tau \}$.
\end{proof}
Theorem 1 indicates that, given the modified value functions $\overline{W}_1(.,\bar{T})$, 
due to the equivalence between $\overline{W}_1(.,\bar{T})$ and $W_1(.)$ in the region 
$\left\{s_0|\overline{W}_1(s_0,\bar{T}) \leq \tau  \right\} $ for any $\tau \in (0,\bar{T})$,
for any $T<\bar{T}$, the minimal reachable set $\mathcal{R}_{\mathrm{min}}(K,T)$ can be characterized by:
\begin{align}
	\label{minreach}
	\begin{split}
	\mathcal{R}_{\mathrm{min}}(K,T)=&\left\{s_0|W_1(s_0)\leq T  \right\}
	=\left\{s_0|\overline{W}_1(s_0,\bar{T})\leq T  \right\}
	\end{split}
\end{align}
Similarly, for any $T<\bar{T}$, the maximal reachable set $\mathcal{R}_{\mathrm{max}}(K,T)$ can be characterized by:
\begin{align}
	\label{maxreach}
	\begin{split}
	\mathcal{R}_{\mathrm{max}}(K,T)=&\left\{s_0|W_2(s_0)\leq T  \right\}
	=\left\{s_0|\overline{W}_2(s_0,\bar{T})\leq T  \right\}
	\end{split}
\end{align}

\begin{remark}
	  {
	In our method, for $T_1,...,T_M \in [0,\infty)$, if $\bar{T}>\max (T_1,...,T_M)$, then the reachable sets 
	of these time horizons can be represented as $T_i-$sublevel sets ($i=1,...,M$) of $\overline{W}_1(.,\bar{T})$ 
	and $\overline{W}_2(.,\bar{T})$, i.e.}
	\begin{align}
		\begin{split}
			&\mathcal{R}_{\mathrm{min}}(K,T_1)=\left\{s_0|\overline{W}_1(s_0,\bar{T})\leq T_1  \right\}\ \ \ \mathcal{R}_{\mathrm{max}}(K,T_1)=\left\{s_0|\overline{W}_2(s_0,\bar{T})\leq T_1  \right\}\\
			&...\\
			&\mathcal{R}_{\mathrm{min}}(K,T_M)=\left\{s_0|\overline{W}_1(s_0,\bar{T})\leq T_M \right\}\ \ \ \mathcal{R}_{\mathrm{max}}(K,T_1)=\left\{s_0|\overline{W}_2(s_0,\bar{T})\leq T_M  \right\}
		\end{split}
	\end{align} 
	  {
	This means that the storage space required to save these reachable sets is independent of $M$.}
\end{remark}

\subsection{Recursive Formulas of the Modified Value Functions}
  {
Value functions $\overline{W}_1(.,.)$ and $\overline{W}_2(.,.)$ are cost-to-go functions, which,
according to Bellman's principle of optimality \cite{dpandhjb}, 
can be expressed as solutions of the following HJ equations with operating cost function $\bar{c}(.)$:
}
\begin{align}
	\label{myHJeq}
	\begin{split}
		\left\{ \begin{array}{l}
			\displaystyle{ \frac{\partial \overline{W}_1}{\partial t}(s,t)=\max_{u\in\mathcal{U}} \left[ \frac{\partial \overline{W}_1}{\partial s}(s,t)\bar{f}(s,u)+\bar{c}(s)  \right] }\\
			\text{s.t. }\overline{W}_1(s,0)=0
		\end{array} \right.\\
		\left\{ \begin{array}{l}
			\displaystyle{ \frac{\partial \overline{W}_2}{\partial t}(s,t)=\min_{u\in\mathcal{U}} \left[ \frac{\partial\overline{W}_2}{\partial s}(s,t)\bar{f}(s,u)+\bar{c}(s)  \right] }\\
			\text{s.t. }\overline{W}_2(s,0)=0
		\end{array} \right.
	\end{split}
\end{align}

  {
According to Eq. (\ref{minreach}) and Eq. (\ref{maxreach}), 
The crucial point in characterizing the minimal and maximal reachable sets is to compute the modified value functions. 
However, $\overline{W}_1(.,.)$ and $\overline{W}_2(.,.)$ are often discontinuous, 
which makes it impossible to obtain the viscosity solutions to the HJ equations in Eq. (\ref{myHJeq}).
In the current research, we use a recursive and interpolation-based approach to approximate these two functions.
}

% For any $\tau\in [0,\bar{T}]$, the following equations hold:
% \begin{align}
% 	\label{recur1}
% 	\begin{split}
% 		\overline{W}^{\bar{T}}_1(s_0)=&\max_{u(.)\in \mathscr{U}} \int_0^{\bar{T}} \bar{c}\left( \bar{\phi}_0^{\bar{T}}(t,s_0,u)  \right) dt 
% 		=\max_{u(.)\in \mathscr{U}} \left[\int_0^{\tau} \bar{c}\left( \bar{\phi}_0^{\bar{T}}(t,s_0,u)  \right) dt \right. \\  
% 		 &+\left.\max_{u'(.)\in \mathscr{U}} \int_\tau^{\bar{T}} \bar{c}\left( \bar{\phi}_0^{\bar{T}}(t,s_0',u')  \right)  \right] \\
% 		=&\max_{u(.)\in \mathscr{U}} \left[\int_0^{\tau} \bar{c}\left( \bar{\phi}_0^{\bar{T}}(t,s_0,u)  \right) dt + 
% 		\overline{W}^{\bar{T}-\tau}_1(s_0') \right]
% 	\end{split}
% \end{align}
% and
% \begin{align}
% 	\label{recur2}
% 	\begin{split}
% 		\overline{W}^{\bar{T}}_2(s_0)=&\min_{u(.)\in \mathscr{U}} \int_0^{\bar{T}} \bar{c}\left( \bar{\phi}_0^{\bar{T}}(t,s_0,u)  \right) dt 
% 		=\min_{u(.)\in \mathscr{U}} \left[\int_0^{\tau} \bar{c}\left( \bar{\phi}_0^{\bar{T}}(t,s_0,u)  \right) dt \right. \\ 
% 		&+ \left.\min_{u'(.)\in \mathscr{U}} \int_\tau^{\bar{T}} \bar{c}\left( \bar{\phi}_0^{\bar{T}}(t,s_0',u')  \right)  \right] \\
% 		=&\min_{u(.)\in \mathscr{U}} \left[\int_0^{\tau} \bar{c}\left( \bar{\phi}_0^{\bar{T}}(t,s_0,u)  \right) dt + 
% 		\overline{W}^{\bar{T}-\tau}_2(s_0') \right]
% 	\end{split}
% \end{align}
% where $s_0'=\bar{\phi}_0^{\bar{T}}(\tau,s_0,u)$.

Denote the discrete form of system (\ref{sys2}) as
\begin{align}
	\label{disdyn}
	s(t+\Delta t)=F(s(t),u(t))
\end{align}
The preceding equation can be yielded by the Euler's method or the Runge-Kutta method.
If $\Delta t$ is small enough, the recursive formula of Eq. (\ref{myHJeq}) is
\begin{align}
	\label{myHJeqdis}
	\begin{split}
		\left\{ \begin{array}{l}
			\displaystyle{  \overline{W}_1(s,(k+1)\Delta t)=\max_{u\in\mathcal{U}} \left[ \overline{W}_1(F(s,u),k\Delta t) + \bar{c}(s)\Delta t  \right] }\\
			\text{s.t. }\overline{W}_1(s,0)=0
		\end{array} \right.\\
		\left\{ \begin{array}{l}
			\displaystyle{  \overline{W}_2(s,(k+1)\Delta t)=\min_{u\in\mathcal{U}} \left[ \overline{W}_2(F(s,u),k\Delta t) + \bar{c}(s)\Delta t  \right] }\\
			\text{s.t. }\overline{W}_2(s,0)=0
		\end{array} \right.
	\end{split}
\end{align}

% For any $\bar{T}\$

\subsection{Approximation of the Modified Value Functions}
\begin{figure}[H]
	\centering
	\subfigure[$s$ lies inside the computational domain.]{\includegraphics[height=0.45\textwidth]{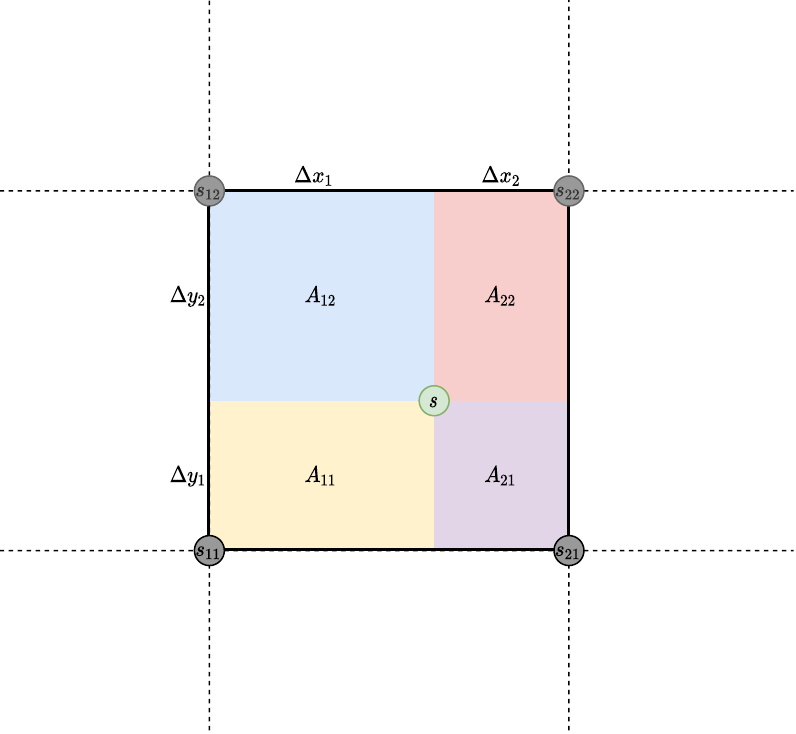}}\ \ \ \ \ \
	\subfigure[$s$ lies outside the computational domain.]{\includegraphics[height=0.45\textwidth]{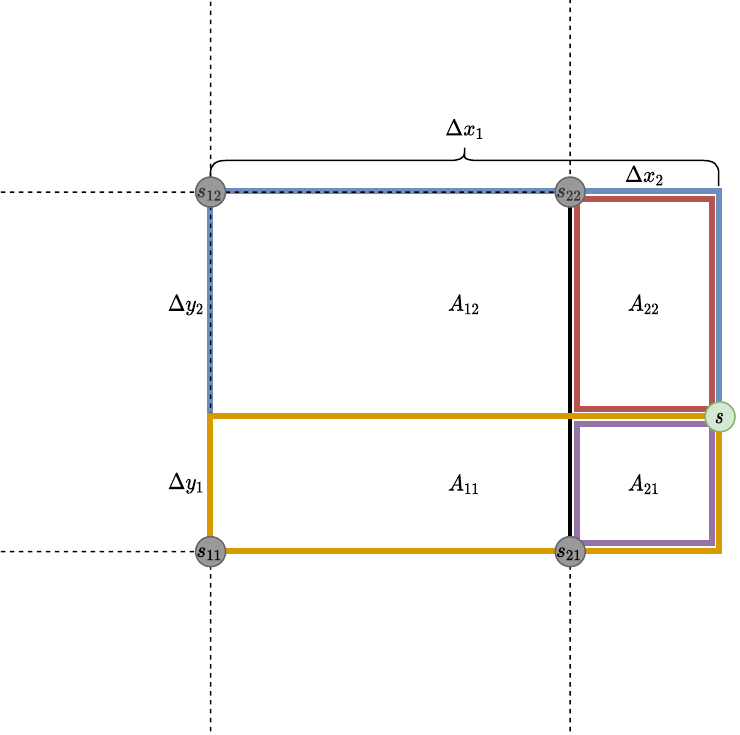}}
	\caption{Approximation of the value function.}
	\label{figchazhi}
\end{figure}

Typically, computing the analytical forms of $\overline{W}_1(.,k\Delta t)$ and $\overline{W}_2(.,k\Delta t)$ for any $k$ 
is difficult. 
In the current study, a rectangular subset of the state space 
is designated as the computational domain, which is divided into a Cartesian grid structure. 
The values of $\overline{W}_1(.,k\Delta t)$ or $\overline{W}_2(.,k\Delta t)$ at each grid point are stored in an array with the same
dimensions as the dynamic system. As an example, consider a two-dimensional system with $s=[x,y]^\mathrm{T}$.  
The function $\overline{W}_1(.,k\Delta t)$ or $\overline{W}_2(.,k\Delta t)$ may be represented by the 
bilinear interpolation of the aforementioned array.
  {
When updating the value of a grid point $s$ near the boundary of the computational domain, 
its transferred state $F(s,u)$ may be outside the computational domain. 
The linear extrapolation is applied to evaluate the value of function at the transferred state. 
% Fig. ***** visualizes the function approximation method used in this paper. 
% Suppose that we want to find the value of the unknown function $W$ at the point $s=[x,y]^\mathrm{T}$. 
As shown in Fig. \ref{figchazhi}, the value of the unknown function $W(.)$ at point $s=[x,y]^\mathrm{T}$, marked by the green point,  
is to be estimated. 
It is easy to find the four nearest grid points to $s$. 
Denote these four grid points as $s_{11}$, $s_{12}$, $s_{21}$, and $s_{22}$.
The value of the function $W(.)$ at $s_{ij}$ is denoted as $v_{ij}$, which is stored in a two-dimensional array in advance,
and denote the coordinate of $s_{ij}$ by $[x_{ij},y_{ij}]^\mathrm{T}$. 
Then $W(s)$ is approximated by 
}
\begin{align}
	\label{appr1}
	W(s)=\frac{v_{11}A_{22} + v_{12}A_{21} +v_{21}A_{12}+v_{22}A_{11}  }{A}
\end{align}
where 
\begin{align}
	\label{appr2}
	\begin{split}
		&A_{ij}=\Delta x_i\Delta y_j \ \forall i,j\in \{1,2\}\\
		&\Delta x_1=x-x_{11}=x-x_{12}, \ \Delta x_2=x_{21}-x=x_{22}-x\\
		&\Delta y_1=y-y_{11}=y-y_{21}, \ \Delta y_2=y_{12}-y=y_{22}-y
	\end{split}
\end{align}
  {
Regardless of whether $s$ is in the computational domain or not, 
the estimation of W is as in Eqs. (\ref{appr1}) and (\ref{appr2}).
$\Delta x_1, \Delta x_2, \Delta y_1, \Delta y_2$ are positive when $s$ lies 
in the computational domain, as shown in Fig. \ref{figchazhi}(a), 
and some of these values are negative when $s$ is outside the computational, as shown in Fig. \ref{figchazhi}(b).
}

Given all the techniques introduced above, the complete
algorithm of the proposed method is described in Algorithm 1. 

\begin{algorithm}[H]
    \caption{Computation of the maximal and minimal reachable sets}
	\begin{algorithmic}[1]
	  \State \textbf{Input:} Dynamic system (\ref{sys1}), admissible control set $\mathcal{U}$, 
	  time horizons $T_1,...,T_M$, target set $K$, number of time steps $m$, a small positive real number $\epsilon$,
	  computational domain $\Omega=[x_l,x_u]\times [y_l,y_u]$, number of grid points $N_x\times N_y$;
	  \State Construct the modified system in discretized form (\ref{disdyn});
	  \State Construct the modified running cost (\ref{runningcost});
	  \State $T_{\max}\leftarrow \max (T_1,...,T_M)$, $\bar{T}\leftarrow T_{\max}+\epsilon$;
	  \State $\Delta t\leftarrow \frac{\bar{T}}{m}$, $\Delta x=\frac{x_u-x_l}{N_x-1}$, $\Delta y=\frac{y_u-y_l}{N_y-1}$;
      \State Let $\mathcal{W}$ and $\mathcal{W}'$ be two $N_x\times N_y$ arrays and initialize them to $\mathbf{0}$; 
	%   \For{$i_x\leftarrow 0,...,N_x-1 $}
	% 	\For{$i_y\leftarrow 0,...,N_y-1 $}
	% 		\For{$i_z\leftarrow 0,...,N_z-1 $}
	% 			  \State $\mathcal{W}[i_x][i_y][i_z]\leftarrow 0$;
    %               \State \begin{align*}
    %                 \mathcal{X}[i_x][i_y][i_z]\leftarrow \left[
    %                     \begin{array}{c}
    %                         x_d+i_x\delta x\\y_d+i_y \delta y\\z_d+i_z \delta z
    %                     \end{array}
    %                 \right]
    %               \end{align*}
	% 		\EndFor
    %     \EndFor
	%   \EndFor
	%   \\

	  \For{$k \leftarrow 1,...,m$}
	  	\State Construct the bilinear interpolation function $\widehat{W}(.)$ using $\mathcal{W}$;
        \For{$i_x\leftarrow 0,...,N_x-1 $}
		  \For{$i_y\leftarrow 0,...,N_y-1 $}
                \State $s_0\leftarrow \left[ x_l+i_x\Delta  x,y_l+i_y \Delta y \right]^\mathrm{T}$; 
		\begin{align*} 
            &\mathcal{W}'[i_x][i_y]\leftarrow
            \begin{cases}
			\max\limits_{u\in \mathcal{U}}  \left[
			\bar{c}\left( s_0 \right) \Delta t + 
			\widehat{W}\left( F(s_0,u)  \right) \right],   \text{for computation of } \overline{W}_1(.,\bar{T})\\
			%%%%%%%%%%%%%%%%%%%%%%%%%%%%%%%%
			\min\limits_{u\in \mathcal{U}}  \left[
			\bar{c}\left( s_0 \right) \Delta t + 
			\widehat{W}\left( F(s_0,u)  \right) \right],  \text{for computation of } \overline{W}_2(.,\bar{T})
			\end{cases}
		\end{align*}
          \EndFor
        \EndFor
        \State Copy $\mathcal{W}'$ to $\mathcal{W}$;
      \EndFor
      \State Return $ \{s|\widehat{W}(s)\leq T_1 \},...,  \{s|\widehat{W}(s)\leq T_M \} $;
    \end{algorithmic}
\end{algorithm}

\subsection{Complexity of the algorithm}
  {
Suppose the number of grid points in each dimension of the computational domain is $N$, 
then the total number of grid points is $N^n$. 
In the level set method, for each grid point, the left and right derivatives in $n$ dimensions need to be
computed. Therefore, at each time step, the time consumed to traverse all grid points is proportional
to $nN^n$. According to the CFL condition (time step size that does not satisfy the CFL condition can
lead to numerical instability), the time step size in the level set method is proportional to the grid size,
which is inversely proportional to $N$. 
Thus, given the time horizon, the required number of time steps is
proportional to $N$. Finally, the time complexity of the level set method is $O(nN^{n+1})$.
}

  {
In our method, for each grid point, the time consumption of the multilinear interpolation is proportional 
to the number of vertices of the $n$-dimensional cube.
Therefore, at each time step, the time consumed to traverse all grid points is proportional to $2^nN^n$.
Since the time step size of our method is not
determined by the CFL condition but is given by the user, the number of time steps is independent of
the number of grids. Finally, the time complexity of our method is $O(2^nN^n)$.
}

  {
The space complexities of our method, and the level set method are the same, 
both require two $n$-dimensional arrays to be stored in memory during the computation. 
Therefore, the space complexities are $O(N^n)$.
}

\section{Two-dimensional system example}
  {
Consider the simple control problem below:}
\begin{align}
    \dot{s}=\frac{d}{dt}\left[\begin{array}{c}
        x\\y
    \end{array}\right]=f(s,u)=\left[\begin{array}{c}
        u\\-x
    \end{array}\right]
\end{align}
  {
In the preceding equation, $s=[x,y]^\mathrm{T}$ is the system state, $u\in\mathcal{U}=[-1,1]$ is the 
control input.
The target set is $K=\{[x,y]^\mathrm{T}| -1< y < 1\}$, and the time horizons are $T_1=0.5$, $T_2=1$, $T_3=1.5$, and $T_4=2$. 
The task is to compute the maximal invariant sets $\mathcal{I}_{\textbf{max}}(K,T_1)$, $\mathcal{I}_{\textbf{max}}(K,T_2)$, 
$\mathcal{I}_{\textbf{max}}(K,T_3)$, and $\mathcal{I}_{\textbf{max}}(K,T_4)$.
This problem can be computed analytically and thereby compared with the results of
our method and the level set method.
According to the duality described by Eq. (\ref{rlt1}), the 
minimal reachable sets of the complement of $K$ can be used to characterize the maximal invariant sets.
}

\subsection{Analytical solution}
  {
The analytical expression of the value function constructed in Eq. (\ref{maxvalue}) with 
$\complement_{\mathbb{R}^n} K$ as the target set is
}
\begin{align}
    W_1(s)=\begin{cases}
        0, &y\leq 1 \lor y\geq 1\\
        -x-\sqrt{-2+x^2+2y}, &-1 <y<1 \land x<0 \land -2 + x^2 + 2 y\geq 0\\
        x-\sqrt{-2+x^2-2y}, &-1 <y<1 \land x\geq 0 \land -2 + x^2 - 2 y\geq 0\\
        \infty, & -1 <y<1 \land x<0 \land -2 + x^2 + 2 y< 0\\
        \infty, & -1 <y<1 \land x \geq 0 \land -2 + x^2 - 2 y< 0
    \end{cases}
\end{align}
  {
where "$\land$" and "$\lor$" are the logical operators "AND" and "OR", respectively.
The analytical expressions of the minimal reachable sets of $\complement_{\mathbb{R}^n} K$ can be obtained from the above equation:
}
\begin{align}
    \begin{split}
        \mathcal{R}_{\text{min}}\left(\complement_{\mathbb{R}^n} K, T \right)
        =&\left\{[x,y]^\mathrm{T}| y\leq -1 \lor y\geq 1 \right\} \cup  \\
        &\left\{[x,y]^\mathrm{T}| -T \leq x <0 \land  -2 + x^2 + 2 y\geq 0 \right\} \cup \\
        &\left\{[x,y]^\mathrm{T}| x\leq -T \land  y\geq xT+ 1+ \frac{1}{2}T^2 \right\} \cup \\
        &\left\{[x,y]^\mathrm{T}| 0 \leq x \leq T \land  -2 + x^2 - 2 y\geq 0 \right\} \cup \\
        &\left\{[x,y]^\mathrm{T}| x\geq T \land  y\leq xT- 1- \frac{1}{2}T^2 \right\}
    \end{split}
\end{align}

% Fig. \ref{figexp11} displays the value function $W_1(.)$ and the maximal invariant sets of this example.

% \begin{figure}[H]
% 	\centering
% 	\includegraphics[width=1.0\textwidth]{fig/figexp11.png}
% 	\caption{Analytical solution of the two-dimensional example. 
% 	The blue surface in the subplot located at the center represents the value function $W_1(.)$, 
% 	and the $T_1, T_2, T_3, T_4$-sublevel sets of this value function characterize the four minimum reachable sets and maximal invariant sets, 
% 	as shown in the four surrounding subplots.}
% 	\label{figexp11}
% \end{figure}

\subsection{Results of the proposed method}
\begin{figure}[H]
	\centering
	\includegraphics[width=1.0\textwidth]{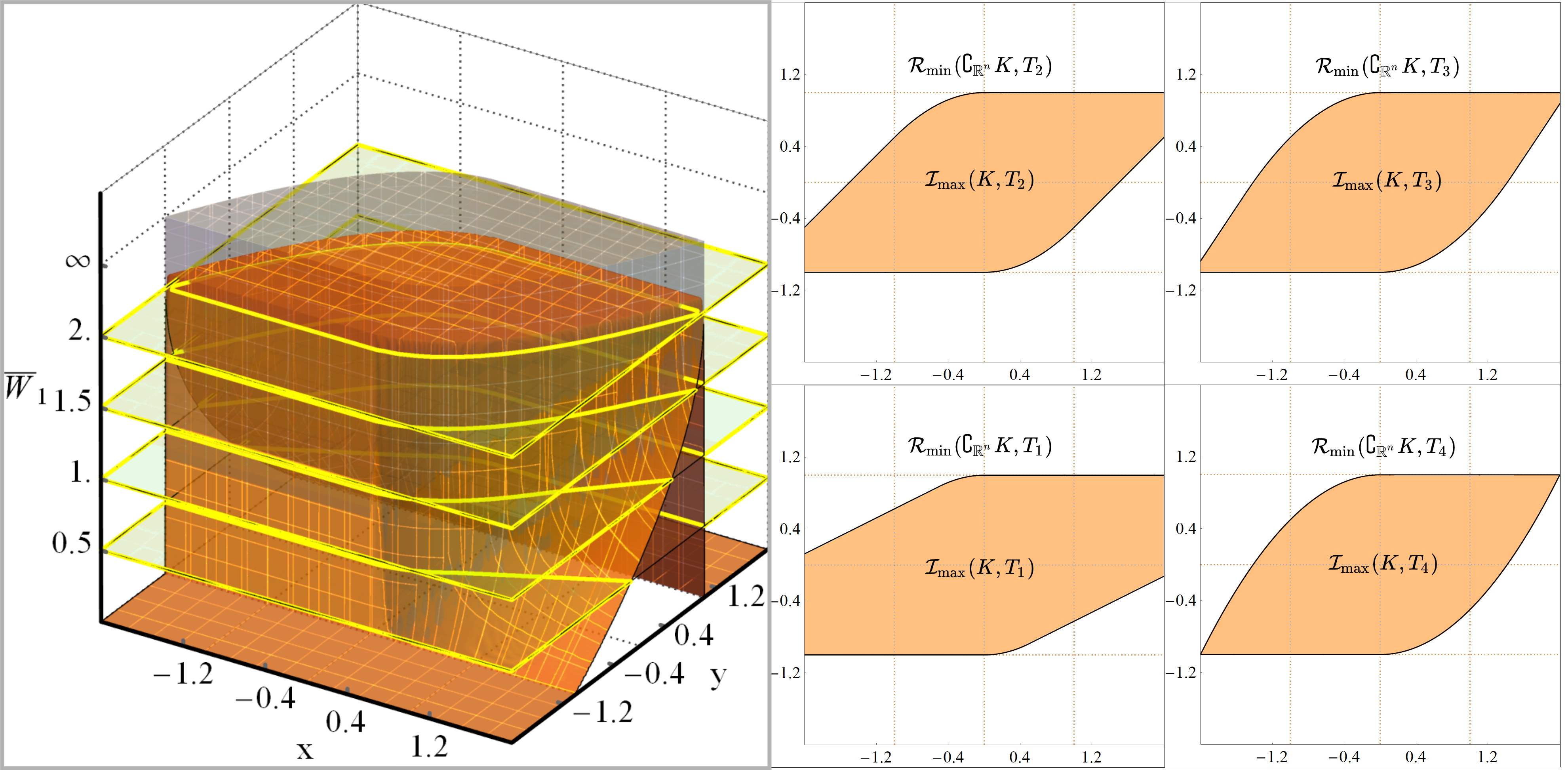}
	\caption{The results of solving the two-dimensional example using our method. 
	In the left subplot, the grey surface 
	represents the analytical solution of $W_1(.)$, 
	and the orange surface represents $\overline{W}_1(.,\bar{T})$ derived from our method. 
	In the four subplots on the right, 
	the orange areas indicate the invariant sets computed by our method, 
	and the analytical solutions of the invariant sets are outlined by the black curves.}
	\label{figexp12}
\end{figure}

  {
We numerically solved this problem using the method described in Algorithm 1 on 
a computational domain $\Omega=[-2,2]\times [-2,2]$ with grid points $201\times 201$. 
The modified system is }
\begin{align}
	\dot{s}= \bar{f}(s,u)=\begin{cases}
		f(s,u), &s\notin \complement_{\mathbb{R}^2} K\\
		\mathbf{0}, &s\in \complement_{\mathbb{R}^2} K
	\end{cases}
\end{align}
and the modified running cost function is 
\begin{align}
	\bar{c}(s)=\begin{cases}
		1, &s\notin \complement_{\mathbb{R}^2} K\\
		0,&s\in \complement_{\mathbb{R}^2} K
	\end{cases}
\end{align}
  {
Since $\max \left(T_1,T_2,T_3,T_4   \right)=2$, $\bar{T}$ is specified as $2.16$ and the time step size $\Delta t$
is set as $0.02$. The 
computational results of our method are shown in Fig. \ref{figexp12}. 
It can be seen that in the region where the function 
$\overline{W}_1(.,\bar{T})$ takes values less than $\bar{T}$, the surfaces representing $W_1(.)$ and 
$\overline{W}_1(.,\bar{T})$ almost overlap,
and the invariant sets computed by the proposed method almost coincide with the analytic solutions.
}

\begin{figure}[H]
	\centering
	\includegraphics[width=0.7\textwidth]{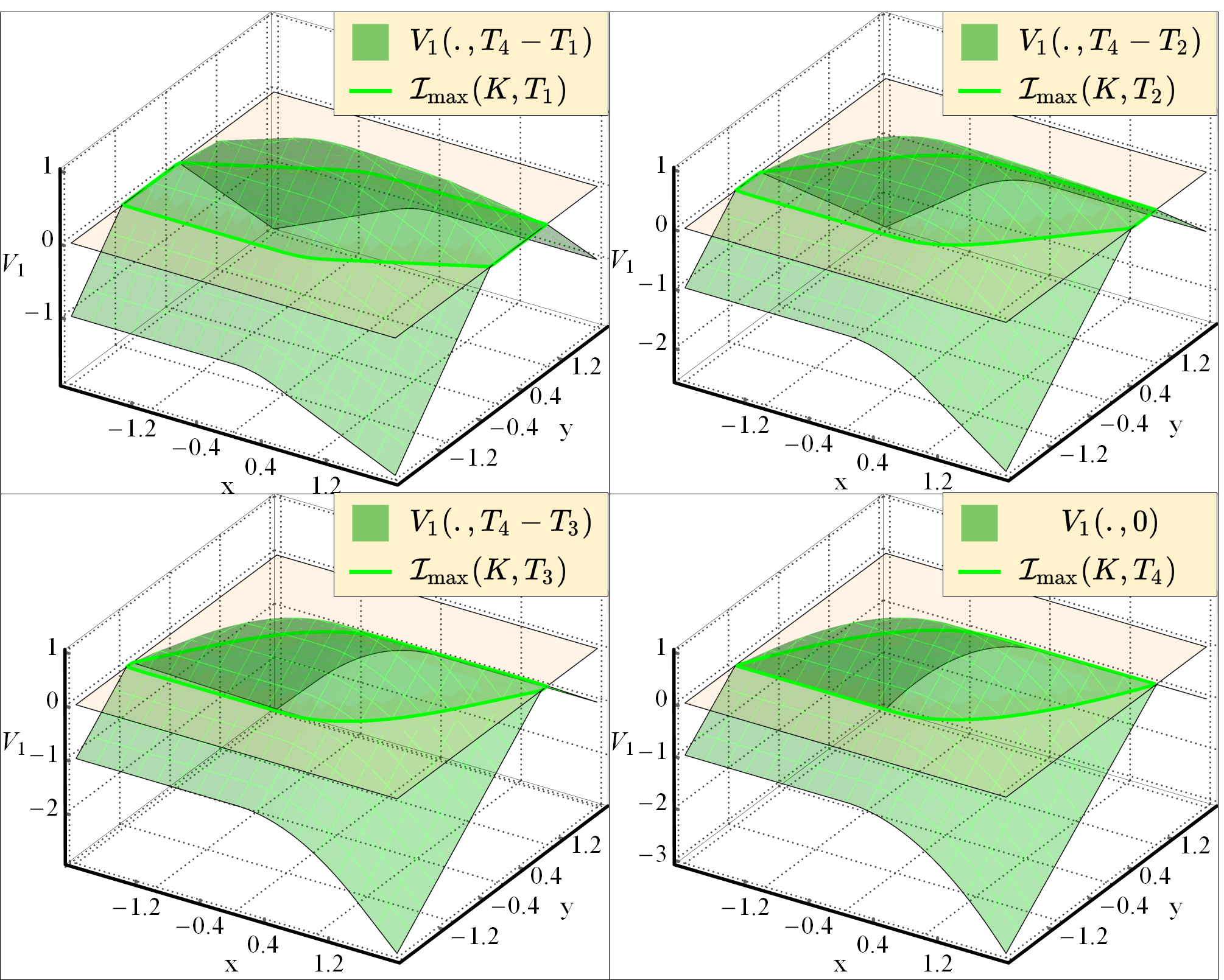}
	\caption{The method of saving invariant sets by the level set method.}
	\label{figexp13}
\end{figure}

  {
Fig. \ref{figexp12} not only displays the results of the proposed method 
but also visually depicts how the results are stored. 
All four invariant sets are represented as the 
complements of the sublevel sets of function $\overline{W}_1(.,\bar{T})$, i.e.
}
\begin{align}
	\begin{split}
		&\mathcal{I}_{\max}(K,T_1)=\complement_{\mathbb{R}^2} \left( \{s|\overline{W}_1(s,\bar{T})\leq T_1 \}   \right)\\
		&...\\
		% \mathcal{I}_{\max}(K,T_2)=\complement_{\mathbb{R}^2} \left( \{s|\overline{W}_1(s,\bar{T})\leq T_2 \}   \right)\\
		% \mathcal{I}_{\max}(K,T_3)=\complement_{\mathbb{R}^2} \left( \{s|\overline{W}_1(s,\bar{T})\leq T_3 \}   \right)\\
		&\mathcal{I}_{\max}(K,T_4)=\complement_{\mathbb{R}^2} \left( \{s|\overline{W}_1(s,\bar{T})\leq T_4 \}   \right)
	\end{split}
\end{align}
  {
This means that only $\overline{W}_1(.,\bar{T})$ needs to be saved to save these four invariant sets.
This is different from the level set method, in which the four invariant sets are represented as zero super-level sets of $V_1(.,.)$ 
at different time if the terminal condition of the HJ equation about $V_1(.,.)$ is set to $V_1(s,T_4)=l(s)$, i.e. 
}
\begin{align}
	\begin{split}
		&\mathcal{I}_{\max}(K,T_1)= \{s|\overline{V}_1(s,T_4-T_1)\geq 0 \}   \\
		&...\\
		% \mathcal{I}_{\max}(K,T_2)=\complement_{\mathbb{R}^2} \left( \{s|\overline{W}_1(s,\bar{T})\leq T_2 \}   \right)\\
		% \mathcal{I}_{\max}(K,T_3)=\complement_{\mathbb{R}^2} \left( \{s|\overline{W}_1(s,\bar{T})\leq T_3 \}   \right)\\
		&\mathcal{I}_{\max}(K,T_4)= \{s|\overline{V}_1(s,0)\geq 0 \} 
	\end{split}
\end{align}
  {
Since $V_1(.,.)$ at different time are usually different from each other, the level set method needs to save these four invariant sets by saving 
$V_1(.,.)$ at four time points, which requiring four times more storage space than our method for the same number of grids. 
The Fig. \ref{figexp13} shows how the level set method saves the invariant sets.}

\subsection{Convergence}
  {
This subsection analyzes the variation of computational error with grid size and time step size. 
The solver settings of Algorithm 1 are listed in Table \ref{tb1}.}
\begin{table}[H]
	\centering
	\caption{Solver settings for Algorithm 1}
	\label{tb1}
	\begin{tabular}{lc}
	\toprule
	\textbf{Parameter} 			  			& 		\textbf{Setting}  \\ \hline
	Computational domain $\Omega$         	&       $[-2,2]\times [-2,2]$     \\   
	Grid points $N_x\times N_y$   			&  		$51\times 51$, $101\times 101$, $151\times 151$, $201\times 201$, $251\times 251$    \\  
	$\bar{T}$               				&      	$2.16$      \\    
	Number of time steps $h$            	&      	$216$, $108$, $72$, $54$    \\ 
	Time step size $\Delta t=\frac{\bar{T}}{h}$     &      $0.01$, $0.02$, $0.03$, $0.04$     \\ \bottomrule
	\end{tabular}
\end{table}
  {
We use the Jaccard index (\cite{Jaccard}) to quantify the errors between the numerical and analytical solutions. 
Specifically, the relative volume error between sets $A$ and $B$ is }
\begin{align}
	e(A,B)=1-\frac{|A\cap B|}{|A\cup B|}
\end{align}

\begin{figure}[H]
	\centering
	\includegraphics[width=0.618\textwidth]{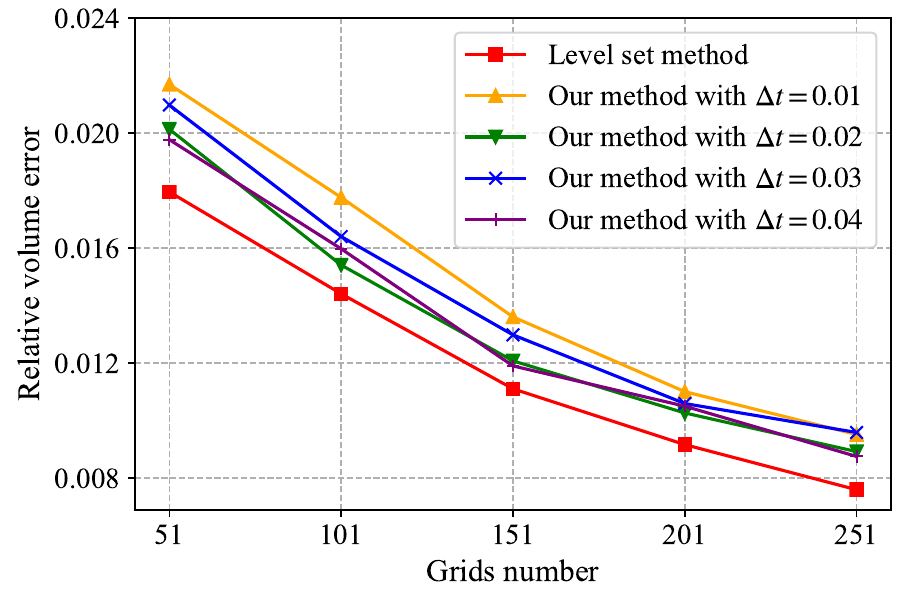}
	\caption{Variations of the relative volume errors with the number of grids.}
	\label{figexp14}
\end{figure}

  {
Take set $\mathcal{I}_{\max}(K,T_4)$ as an example, Fig. \ref{figexp14} shows the relative volume errors under different time step sizes against the number of grid points per dimension.
The level set method is also involved in the comparison.
It can be seen that the computational accuracy of the proposed method is 
not sensitive to the time step size and is not significantly different from that of the level set method.}

% our method needs to compute the minimal reachable set of the complement of $K$.

\section{Aircraft ground motion example}
\begin{figure}[H]
	\centering
	\includegraphics[width=8cm]{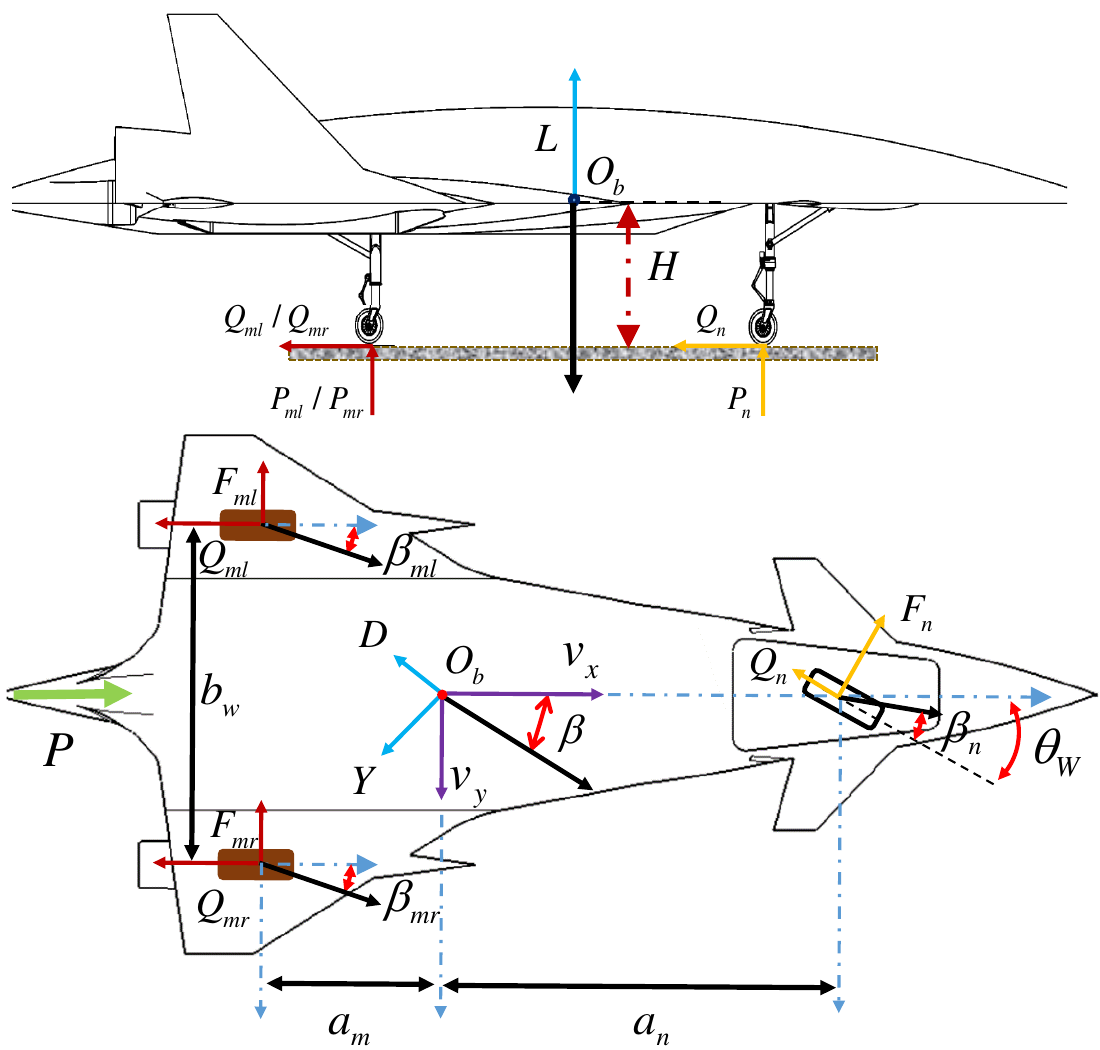}
	\caption{Aircraft ground motion}
	\label{figground}
\end{figure}

\begin{figure}[H]
	\centering
	\includegraphics[width=8.85cm]{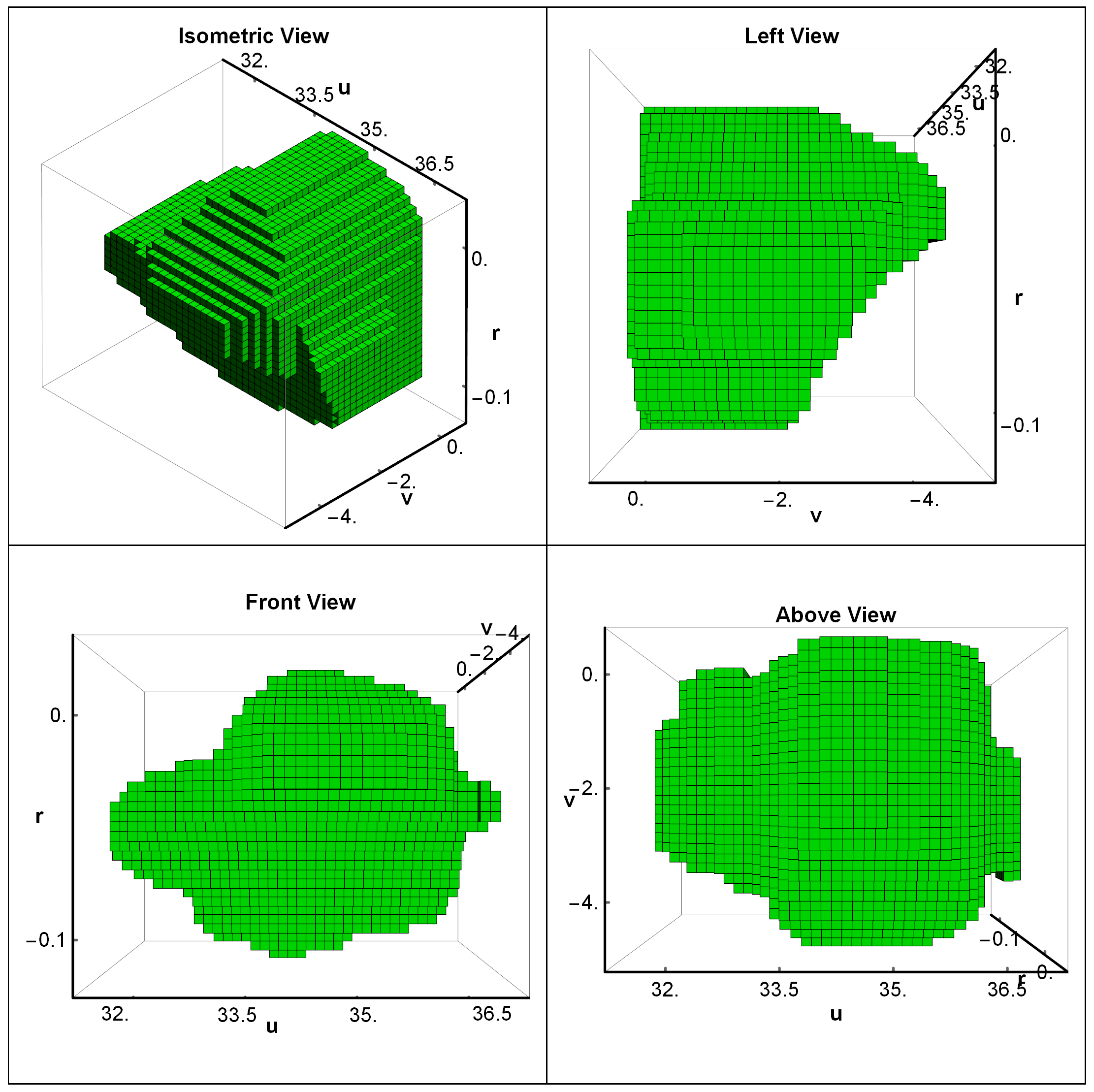}
	\caption{Irregular target set}
	\label{figirr}
\end{figure}

Consider a simplified aircraft ground motion dynamics system:
\begin{align}
	\dot{s}=\left[
		\begin{array}{c}
			\dot{v}_x\\\dot{v}_y\\\dot{r}
		\end{array}
	\right]=
	\left[\begin{array}{c}
		{rv_y+\frac{F_x}{m}}\\
		{-rv_x+\frac{F_y}{m}}\\
		{\frac{M_z}{I_{z}}}
	\end{array}\right]
\end{align}
where $s=[v_x,v_y,r]^\mathrm{T}$ is system state.
$v_x$ and $v_y$ are the longitudinal and lateral velocities, respectively, $r$ is the yaw rate,
$F_x$ and $F_y$ are the longitudinal and lateral resultant forces, $M_z$ is the resultant moment. 
The expressions for $F_x,F_y,M_z$ are as follows:
\begin{align}
	\begin{cases}
		F_x=-D \cos \beta - Y \sin \beta -Q_n-Q_{ ml }-Q_{mr}+P\\
		F_y=Y \cos \beta -D\sin\beta-F_n-F_{ml}-F_{mr}\\
		\displaystyle{M_z=n+\left(F_{ml}+F_{mr} \right)a_m-F_n a_n +\left(  Q_{ ml }+Q_{mr} \right)\frac{b_w}{2}}
	\end{cases}
\end{align}
where $D=\frac{1}{2} \rho v^2 S C_{Y\beta} \beta$ is the drag, $Y=\frac{1}{2} \rho v^2 S C_{D0}$ is the aerodynamic lateral force,
$F_n,F_{ml},F_{mr}$ are the ground lateral forces, $Q_n,Q_{ ml },Q_{mr}$ are the ground longitudinal forces. 
See Figure. \ref{figground} for the specific meanings of these variables.
Assume that the ground friction coefficient $\mu=0.04$, then
these variables can be derived by solving the force and moment balance equations (\cite{a0016}).
It is worth pointing out that the expression of the lateral force of the nose-wheel is:
\begin{align}
	\label{eqfn}
	\begin{split}
	F_n=-\mu_d \sin \left\{ \mu_c \arctan \left(\mu_b \tan \beta_n\right) \right. - 
	 \left. \mu_e \left[\mu_b \tan \beta_n-\arctan \left(\mu_b \tan \beta_n\right)  \right] \right\}P_n
	\end{split}
\end{align}
and
\begin{align}
	\label{eqbn}
	\beta_n=\arctan \frac{ (v+ra_n)\cos \theta_W-u\sin \theta_W  }{ u\cos \theta_W+(v+ra_n)\sin \theta_W  }
\end{align}
where $\theta_W\in [-0.15,0.15]$ is the nose-wheel deflection and is considered as the control input.
See literatures (\cite{a0017,a0018}) for the meaning of each symbol in Eq. (\ref{eqfn}) and Eq. (\ref{eqbn}).
In summary, the aircraft ground motion is a highly nonlinear system, that it cannot be reduced into an affine nonlinear form.
Moreover, in this example, we let the target set $K$ be a set with irregular shape,
which is regarded as a combination of some cubic cells,
and whether a cell is
contained in the target set is determined by its center, see Figure. \ref{figirr}. 
Due to these factors, this example cannot be solved by level set method.

The parameters of the reference aircraft are summarized in Table \ref{table2}.
\begin{table}[H]
    \centering
    \caption{Numerical values representative of the reference aircraft}
    \label{table2}
    \begin{tabular}{ll|ll}
    \hline
    \textbf{Parameter}    & \textbf{Value}    &  \textbf{Parameter}                & \textbf{Value} \\ \hline
    $m$       &  104915.9           &  $I_{z}$   &  10504308.1  \\
    $S$   &  249.9             &  $\bar{b}$                    &  12.1 \\
    $\rho$     &  1.293        &  ${b}_w$                  &  23.8 \\
    $a_n$          &  17.9         &       $a_m$       &  2.3   \\
    $b_W$      &  6.9        &  $\mu_d$       &  0.1014 \\
    $\mu_b$         &  -10.11         &  $\mu_c$      &  1.438  \\ 
	$\mu_e$           &  -0.8507         &  $C_{D0}$    &  0.061 \\
    $C_{Y\beta}$      &  -1.4            &  $C_{n\beta}$        &  0.2 \\
	$C_{nr}$        &  -1.5         &   $H$    &  9.47 \\ 
	$C_{L0}$        &  -0.053         &        &    \\\hline
    \end{tabular}
\end{table}

\begin{figure}[H]
	\centering
	\includegraphics[width=8.85cm]{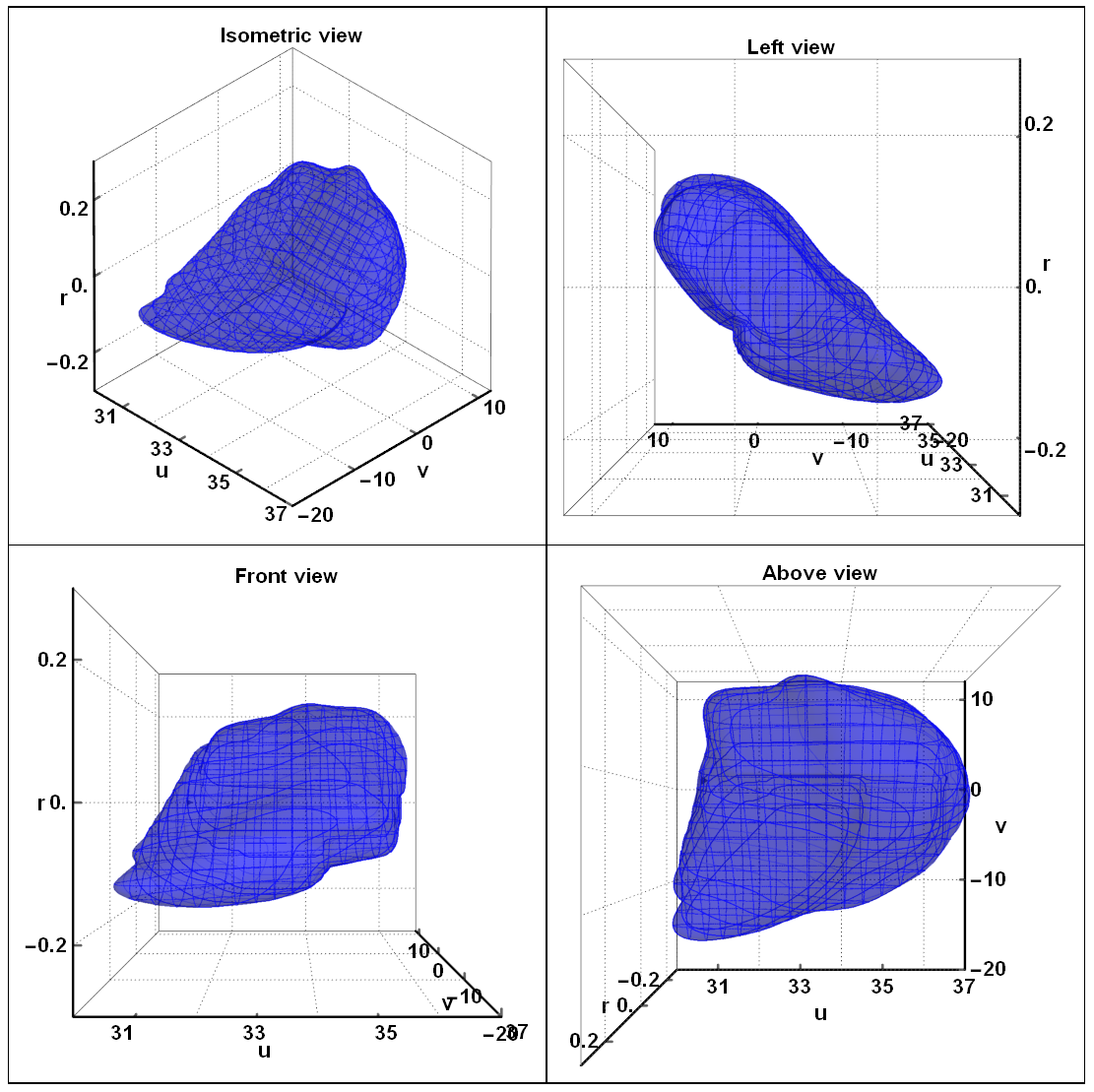}
	\caption{Computation result of the maximal reachable set.}
	\label{figexp3}
\end{figure}

The control objective is to steer the state towards the target set within time horizon $T=2$.
In other words, the maximal reachable set $\mathcal{R}_{\max}(K,T)$ needs to be computed.
Table \ref{tb2} summarizes the parameters of Algorithm 1.
\begin{table}[H]
	\centering
	\caption{Solver settings for Algorithm 1}
	\label{tb2}
	\begin{tabular}{lc}
	\toprule
	\textbf{Parameter} 			  			& 		\textbf{Setting}  \\ \hline
	Computational domain $\Omega$         	&       $[30,37]\times [-20,12]\times [-0.3,0.3]$     \\   
	Grids					&  		$101\times 101\times 101$    \\  
	$\bar{T}$               				&      	$2.1$      \\    
	Number of time steps $h$            	&      	$210$   \\ 
	Time step size $\Delta t=\frac{\bar{T}}{h}$     &      $0.01$     \\ \bottomrule
	\end{tabular}
\end{table}
Figure \ref{figexp3} depicts the maximal reachable set.
% \begin{align}
% 	\mathcal{R}_{\text{max}}(K,T)=\left\{s_0| \overline{W}_2^{201\Delta t}(s_0)\leq 2 \right\}
% \end{align}
% a $2$ level set of the value function $\overline{W}_2^{201\Delta t}(.)$,
% See Figure. ***.

% \subsection{Viable set}
% In this example, we aim to keep the aircraft state in the target set $K=[22,48]\times[-18,18] \times[-0.4,0.4]$
% for more than the given time horizon $T=2$. The solver settings are the same as those in the previous example.
% The computation result is shown in Figure. \ref{figexp4}.
% \begin{figure}[H]
% 	\centering
% 	\includegraphics[width=8.85cm]{fig/figexp4.png}
% 	\caption{Computation result of the viable set.}
% 	\label{figexp4}
% \end{figure}

\section{Conclusions}
This paper proposes a unified framework for dealing with reachability and invariance problems. In this framework, 
the reachable or invariant sets with different time horizons are characterized by a family of non-zero sublevel sets 
of the solution of an HJ PDE with a running cost function, which is approximated by recursion and interpolation. 
This mechanism avoids the computation of the dissipation function and can reduce storage space consumption compared to the level set method. 
It also avoids the construction of the signed distance function of the target set, 
which acts as the terminal condition of the HJ PDE, such that it can handle the irregular target sets.

The suggested method, like many others, suffers from the exponential escalation 
in memory and computational cost as the system's dimension grows (\cite{a008,i7}). 
To overcome this problem, some ideas are worth taking into account,
such as 
decomposing a high-dimensional system into a number of lower-dimensional subsystems 
based on dependencies (\cite{a002,i9,ijc1}) or differences in the rate of change (\cite{i12}) between state variables,
and solving these sub-problems sequentially.

Improving the interpolation algorithm to reduce the required quantity of 
grids might be another effective approach to overcome the curse of dimensionality.
  { 
Furthermore, in our method, different running cost functions can be constructed for 
real problems, such as fuel consumption, distance traveled per unit of time. 
Then some more generalized reachability 
or invariance problems can be discussed. These will be considered in our future work.
}

\bibliographystyle{unsrt}
% \bibliography{Bibliography/IEEEabrv,Bibliography/BIB_xx-TIE-xxxx}\ %IEEEabrv instead of IEEEfull
\bibliography{mybib}

\end{document}